\documentclass[aip,amsmath,amssymb,
 reprint,%
]{revtex4-1}

\usepackage{graphicx}
\usepackage{dcolumn}
\usepackage{bm}
\usepackage[colorlinks=true, linkcolor=blue, citecolor=blue, urlcolor=blue]{hyperref}

\usepackage[utf8]{inputenc}
\usepackage[T1]{fontenc}
\usepackage{mathptmx}
\usepackage{etoolbox}
\usepackage{multirow}
\usepackage{siunitx}
\usepackage{xcolor}
\usepackage{multirow}

\makeatletter
\def\@email#1#2{%
 \endgroup
 \patchcmd{\titleblock@produce}
  {\frontmatter@RRAPformat}
  {\frontmatter@RRAPformat{\produce@RRAP{*#1\href{mailto:#2}{#2}}}\frontmatter@RRAPformat}
  {}{}
}%
\makeatother
\begin{document}

\preprint{AIP/123-QED}

\title{First-principles study of metal--biphenylene interfaces: 
    structural, electronic, and catalytic properties}

\author{Maicon P. Lebre}
\affiliation{Departamento de F\'isica, Universidade Federal de Lavras, C.P. 3037, 37203-202, Lavras, MG, Brazil}
\author{Dominike Pacine}
\affiliation{Instituto Federal de Educa\c{c}\~ao, Ci\^encia e Tecnologia do Tri\^angulo Mineiro, Campus Uberlândia, C.P. 1020, 38064-190, MG, Brazil}
\author{Erika N. Lima} 
\affiliation{Instituto de Física, Universidade Federal de Mato Grosso, 78060-900, Cuiabá, MT, Brazil}
\author{Alexandre A. C. Cotta}
\author{Igor S. S. de Oliveira}
    \email{igor.oliveira@ufla.br}
\affiliation{Departamento de F\'isica, Universidade Federal de Lavras, C.P. 3037, 37203-202, Lavras, MG, Brazil}

\date{\today}

\begin{abstract}

We employ first-principles density functional theory (DFT) calculations to investigate the structural, electronic, and catalytic properties of biphenylene supported on various metal substrates. The substrates considered are the (111) surfaces of Ag, Au, 
Ni, Pd, Pt, Cu, Al, and the Cu$_3$Au alloy. Our results reveal how the interaction between biphenylene and the substrate governs its stability, degree of corrugation, electronic hybridization, and interfacial charge transfer. In particular, we observe a clear trend where weakly interacting metals preserve the intrinsic features of biphenylene, while more reactive substrates lead to significant structural and electronic modifications. We further evaluate the hydrogen evolution reaction (HER) activity of these systems, showing that certain metal supports, especially Pd, Pt, Ag, and Cu, can enhance the catalytic performance of biphenylene. Notably, Ag and Cu combine good catalytic activity with lower cost and chemical stability, offering a promising balance for practical applications. These findings provide insights into the design of biphenylene–metal interfaces, supporting their use in next-generation electronic and catalytic devices.
 
\end{abstract}

\maketitle

\section{\label{sec:level1} Introduction}

Two-dimensional (2D) materials have received significant interest in recent years due to their exceptional electronic, mechanical, and optoelectronic properties, which arise from their atomically thin structures and quantum confinement effects. These unique characteristics make them highly promising candidates for a wide range of technological applications, including next-generation transistors, flexible electronics, energy storage devices, and photodetectors.
\cite{liu2019recent, shanmugam2022review}
Among these materials, graphene, a single layer of carbon atoms arranged in a hexagonal lattice, has emerged as the most well-studied 2D system, exhibiting remarkable electrical conductivity, mechanical strength, and thermal stability. 
\cite{novoselov2004,geim2007}
The success of graphene has not only revolutionized materials science but also initiated the exploration of alternative 2D carbon-based nanostructures, such as graphyne and graphdiyne, which exhibit distinct structural and electronic properties.
\cite{jana2021emerging}
Beyond pure carbon systems, the broader family of 2D materials, including transition metal dichalcogenides (TMDs) \cite{naclerio2023review} and hexagonal boron nitride (h-BN), \cite{manzeli20172d} has further expanded the possibilities for designing novel devices with tailored functionalities. 
\cite{bhimanapati2015recent,lei2022graphene}
As research progresses, the precise engineering of these materials continues to unlock new opportunities in nanotechnology and quantum materials research.

Recently, biphenylene (BPN), a novel 2D carbon allotrope, has attracted significant attention due to its unique atomic architecture, which consists of a non-benzenoid network of four-, six-, and eight-membered carbon rings. \cite{fan2021biphenylene} This distinctive arrangement endows BPN with electronic and mechanical properties that markedly differ from those of graphene and other 
$sp^2$--hybridized carbon frameworks, \cite{biphenelene-sci} offering new opportunities for tailoring material properties at the atomic level.
Notably, BPN exhibits a metallic character even in
narrow ribbons, in contrast to graphene. Its high thermal stability (up to 4500 K) and
robust mechanical properties (Young’s modulus $\approx 260$~N/m) \cite{Luo2021} make it a
strong candidate for different advanced applications.

The integration of 2D materials with metallic substrates plays a crucial role in their potential applications, particularly in electronics, \cite{radisavljevic2011single} catalysis, \cite{wang2018catalysis}, and sensing devices.
\cite{he2012graphene,late2013sensing}
The choice of substrate is especially important from an experimental physics perspective, as it can strongly influence the growth, stability, and properties of the 2D material.
For BPN, as with other 2D materials, \cite{giovannetti2008doping,khomyakov2009first,cai2019integration} the interaction with the metal substrate can induce significant modifications in its electronic structure, such as charge transfer, band alignment, and hybridization effects, which in turn affect its transport, adsorption, catalytic behavior, and chemical stability. Such interactions are not merely perturbative but can be harnessed to deliberately tailor the properties of BPN for targeted applications.
The choice of metallic substrate is thus of critical importance. Substrates that interact weakly, primarily via van der Waals forces, help preserve the intrinsic properties of BPN, while strongly interacting substrates can induce substantial modifications such as the opening of band gaps. 

Furthermore, the decoration or interfacing of BPN with transition metals (such as Cu, Pd, Co, and others) has been shown to alter its electronic and adsorption properties, as well as its catalytic activity. \cite{CLuo2024,Lakshmy2023} 
The nature of the metal–carbon bond and the extent of charge transfer are critical factors that determine the suitability of a given substrate for a targeted application, such as hydrogen evolution reaction (HER) catalysis or gas sensing. These findings underscore the importance of substrate selection and interface engineering, not only for fundamental studies but also for guiding experimental efforts toward the realization of high-performance devices.

Moreover, 2D materials are generally synthesized by either top-down or bottom-up methods.\cite{dong2018interface} 
Bottom-up approaches use atoms or molecules as precursors and can produce large-area, homogeneous layers of 2D materials on a substrate via methods such as chemical vapor deposition (CVD), molecular beam epitaxy (MBE), or wet chemical synthesis. Given their potential for scalable fabrication, it is of great interest to explore the growth of BPN on metal substrates and understand the underlying substrate–material interactions. In fact, recent experiments have demonstrated the synthesis of BPN on Au(111)
\cite{fan2021biphenylene} and the formation of BPN dimers on Ag(111) \cite{zeng2021chemisorption} surfaces, underscoring the relevance of metal substrates in guiding its growth. Finally, these results motivate a deeper theoretical investigation into how different metal(111) surfaces influence the structural and electronic properties of BPN.

In this work, we employ first-principles density functional theory (DFT) simulations to systematically investigate the structural, electronic, and bonding properties of BPN interfaced with various metal(111) surfaces, namely, Ni, Pd, Pt, Al, Cu, Ag, Au, and Cu$_3$Au. The chosen metal(111) surfaces span a diverse range of chemical reactivity, electronic configuration, and catalytic performance. From reactive transition metals (Ni, Pd, Pt) to more inert noble metals (Cu, Ag, Au), as well as a $p$-block metal (Al) and a model alloy (Cu$_3$Au). This selection allows us to systematically explore how the nature of the substrate influences the structural, electronic, and catalytic properties of BPN and to draw trends across different bonding regimes. We analyze the adsorption configurations, stability, charge transfer characteristics, and metal-induced modifications in the electronic structure of BPN. Additionally, we explore the implications of these interactions for potential applications in nanoelectronics and catalysis, particularly in the hydrogen evolution reaction (HER). Our findings provide fundamental insights into the behavior of BPN at metal interfaces, contributing to the broader understanding of metal–2D material interactions and their technological applications.

\section{Computational details}

The calculations were performed based on the density-functional theory (DFT), 
as implemented in the Vienna \textit{ab initio} simulation package (VASP).
\cite{VASP} The generalized gradient approximation (GGA), in the form
proposed by Perdew, Burke and Ernzerhof, \cite{PBE} was used to describe the
exchange--correlation functional. The Kohn--Sham orbitals were expanded in a plane
wave basis set with an energy cutoff of 400~eV. The electron--ion interactions
were taken into account using the Projector Augmented Wave (PAW) method. \cite{Kresse99} 
All geometries have been relaxed until atomic forces were lower than 25~meV/\AA. The Brillouin Zone was sampled according to the
Monkhorst--Pack (MP) method,\cite{Monkhorst} using a 12$\times$12$\times$12 mesh for geometry optimizations of bulk metals, 12$\times$12$\times$1 for
BPN,
and a 4$\times$8$\times$1 $k$-points grid to simulate the BPN/metal(111) structures. A large vacuum region between 15 to 20~\AA\ was used in the direction perpendicular to the structures, avoiding the interaction between periodic images. 
The effects of van der Waals forces are important to describe the interactions between BPN and metallic substrates, thus we have also included a functional accounting for dispersion effects, representing
those interactions, according to the method proposed by Grimme (DFT-D3). 
\cite{grimmeD3}

\section{Results and Discussion}

\subsection{Structural properties}

\begin{figure}
    \centering
    \includegraphics[width=0.95\linewidth]{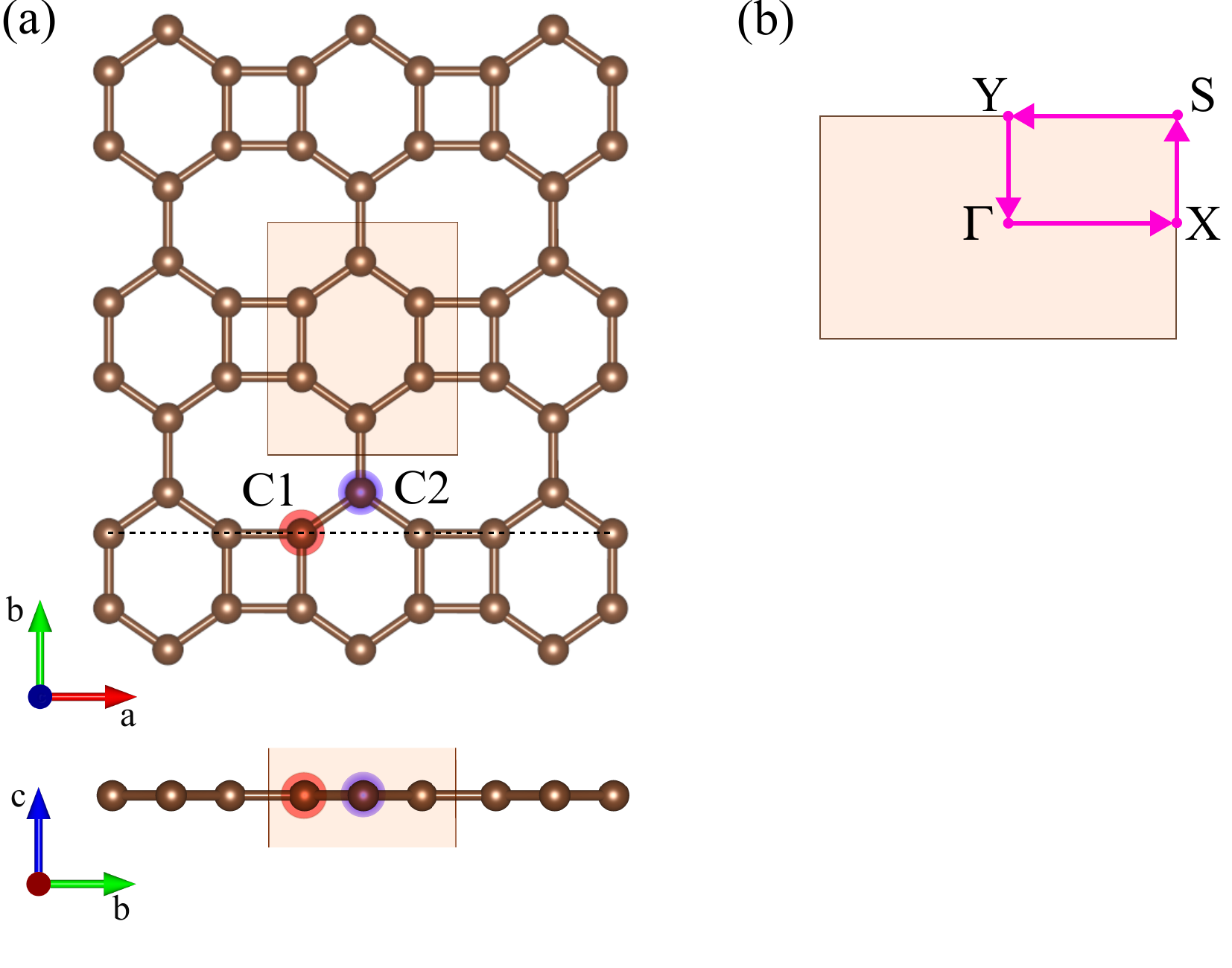}
    \caption{The upper and lower panels show the top and side views of biphenylene, respectively. The orange shaded region indicates the unit cell of the biphenylene crystal lattice. Red and blue spheres, labeled C$_1$ and C$_2$, represent the adsorption sites considered in the hydrogen evolution reaction (HER) study. The dashed black line marks the direction parallel to the plane analyzed in the ELF (Electron Localization Function) calculations. (b) Reciprocal lattice of biphenylene, indicating the high-symmetry points used in the electronic band structure calculations.}
    \label{fig:BFN}
\end{figure}

The BPN unit cell (Fig.~\ref{fig:BFN}) presents a rectangular geometry, where we find
the lattice parameters $a_{\rm BPN}=3.76$~\AA\ and $b_{\rm BPN}=4.52$~\AA, 
in good agreement with previous DFT simulations. \cite{Luo2021}
The lattice constants ($a_{\rm met}$) for the eight bulk metals considered in 
this work are listed in Table~\ref{tab:strain},  
showing a concordance with experimental values within less than 2\%. 
\cite{lu1992electronic, okamoto2016crystal}

\begin{figure*}
    \includegraphics[width=\textwidth]{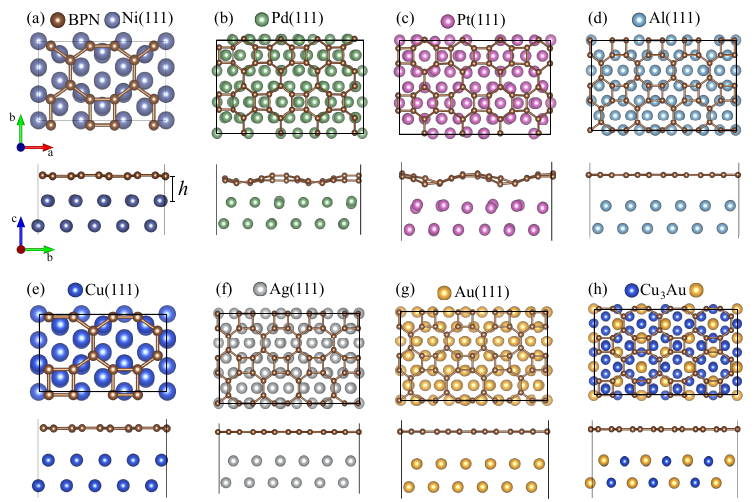}
    \caption{\label{fig:struc}  Top and side views illustrate the most energetically favorable configurations of BPN adsorbed on the metal(111) surfaces examined in this work, after structural relaxation. The configurations correspond to (a) BPN/Ni(111), (b) BPN/Pd(111), (c) BPN/Pt(111), (d) BPN/Al(111), (e) BPN/Cu(111), (f) BPN/Ag(111), (g) BPN/Au(111), and (h) BPN/Cu$_3$Au(111). The black solid lines outline the unit cells of the crystalline structures in each system. The variable \textit{h} represents the vertical separation between the BPN and the underlying metal substrate.}
\end{figure*}

In the seminal work of Fan et al., \cite{fan2021biphenylene} a bottom-up growth of BPN is performed on top of Au(111) surface, while in Ref.~\onlinecite{zeng2021chemisorption} the formation of BPN dimers occurs on top of Ag(111).  
Motivated by these experimental works and considering that (111) surfaces are the most stable and closely packed facets of face-centered cubic (FCC) metals, we chose to construct various metal(111) surfaces (Ni, Pd, Pt, Al, Cu, Ag, Au) to systematically investigate the deposition and interaction of BPN. We also included the Cu$_3$Au(111) alloy surface, which combines the weak physisorption characteristics of noble metals with enhanced thermal stability. Previous studies have demonstrated that Cu$_3$Au(111) enables the successful on-surface synthesis of graphene nanostructures without strong hybridization or distortion.
\cite{cahlik2023versatile}
The FCC structure of the bulk metals presents (111) surfaces with lattice constant
$a_s = a_{\rm met}/\sqrt{2}$. To reduce the lattice mismatch between BPN and
substrate, we use the periodicity listed in Table~\ref{tab:strain} for BPN
and metals(111). The chosen parameters ensure that the average strain 
($\overline{\epsilon}$, see Ref.~\onlinecite{straineq}) in the supercell is at maximum $\sim 5\%$.
Since the band structure of BPN is sensitive to applied strains, 
\cite{hou2023opening, kuritza2024dir}
we choose to keep the lattice parameters of BPN and stretch/compress the metallic
substrate. 

As will be discussed in Section~\ref{sec:en}, we consider various 
deposition sites for BPN on the metals(111) surface; these configurations are represented in Fig.~S-1 of the Supplementary Material. The interfaces were constructed by placing a layer of BPN on top of a metal substrate composed of six atomic layers. During structural relaxation, the bottom two layers were kept fixed.

In Fig.~\ref{fig:struc} we show the energetically most stable configuration for each system, from both top and front views. 
We observe that after relaxation, the metal surface and BPN planar structure
are mostly preserved, although for some surfaces, such as Pd(111) and Pt(111), the
BPN clearly becomes corrugated, which is caused by the interface interactions. 
To quantify the degree of corrugation we introduce an average out-of-plane distortion parameter ($\xi$), \cite{han2016possible} written as
\begin{equation}
    \xi = \sqrt{\frac{\sum\limits_{i=1}^{N_C}(h_i - \bar{h})^2}{N_c - 1}},
\end{equation}
where $N_C$ is the number of carbon atoms in the supercell, $h_i$ is the distance
of each C-atom to the metal surface, and $\bar{h}$ the average distance between
BPN atoms to the metal surface. 
The results are presented in Table~\ref{tab:strain}, where we observe that BPN on the Pd(111) and Pt(111) present $\xi$ values an order of magnitude larger than the other surfaces. The interaction with BPN also slightly distorts the topmost metal layer, although the overall planarity of the substrate is preserved. This structural response highlights how different metal substrates induce varying degrees of distortion in BPN, which can influence its electronic and catalytic properties. 

\begin{table*}
\caption{\label{tab:strain} Optimized bulk metals lattice parameters ($a_{\rm met}$), BPN ($m\times n$) and metal ($p\times q$)  supercell periodicity, average lattice mismatch ($\bar{\epsilon}$), average BPN--metal(111) interlayer distance ($\bar{h}$), out-of-plane BPN corrugation parameter ($\xi$), adsorption energy ($E_{\rm ads}$), formation energy ($E_{\rm for}$), charge transfer ($\Delta\rho$), and work function (WF) for various BPN/metal(111) systems.}
\begin{ruledtabular}
\begin{tabular}{ccccccccccc}
    metal(111)     & bulk      & biphenylene & substrate  &  mismatch & $\bar{h}$ 
    & $\xi$ & {$E_{\rm ads}$} & $E_{\rm for}$ & $\Delta\rho$ & WF \\
    substrate & $a_{\rm met}$~(\AA) & ($m\times n$)   & ($p\times q$) & $\overline{\epsilon}$ (\%) & (\AA) 
    & (\AA) & {(meV/C)} & (eV/C) & ($10^{13}e\cdot\rm{cm}^{-2}$)  & (eV)  \\
    \hline
    Ni & 3.51 & ($2\times 1$) &  ($3\times\sqrt{3}$)  & 2.74	 
    & 1.95 & 0.04 & 557.6  & 0.031 & 28.94 & 4.95  \\
    Pd & 3.94 & ($4\times 2$) & ($3\sqrt{3}\times 3$)  & 5.50	
    & 2.33 & 0.25 & 172.9  & 0.353 & 10.36 & 4.88   \\
    Pt & 3.97 & ($4\times 2$) & ($3\sqrt{3}\times 3$)  & 4.38	 
    & 2.59 & 0.30 & 157.1  & 0.369 & 5.56 & 5.56  \\
    Al & 4.04 & ($4\times 2$) & ($3\sqrt{3}\times 3$)  & 3.28	
    & 3.10 & 0.02 & -225.9 & 0.751 & 10.90 & 4.28  \\
    Cu & 3.63 & ($2\times 1$) &  ($3\times\sqrt{3}$)  & 2.00	 
    & 2.52 & 0.03 & 113.2  & 0.413 & 13.53 & 4.92  \\
    Ag & 4.15 & ($4\times 2$) & ($3\sqrt{3}\times 3$)  & 2.03 
    & 3.05 & 0.01 & 83.1  & 0.443 & 6.64 & 4.35   \\
    Au & 4.16 & ($4\times 2$) & ($3\sqrt{3}\times 3$)  & 2.02 
    & 3.26 & 0.01 & 78.7  & 0.447 & -0.40 & 5.12 \\
    Cu$_3$Au & 3.78 & ($4\times 2$) & ($6\times 2\sqrt{3}$) & 4.19  
    & 3.23 & 0.01 & 87.1 & 0.439 & 1.42 & 4.87  \\
\end{tabular}
\end{ruledtabular}
\end{table*}

\subsection{\label{sec:en} Energetic stability}

To assess the stability of BPN on metal surfaces, we evaluate its adsorption energy ($E_{\rm ads}$) and formation energy ($E_{\rm for}$).
These energetic parameters provide insight into the strength of interaction and thermodynamic feasibility of BPN/metal(111) interfaces.
Firstly, we analyze different BPN deposition configurations on the metal(111) substrate, 
as represented in Fig.~S-1 of the Supplementary Material. 
For each one, the adsorption energy per C-atom is computed according to 
\begin{equation}
    \label{eq:Eads}
    E_{\rm ads} = (E_{\rm BPN} + E_{\rm sub} - E_{\rm BPN/sub})/N_C,
\end{equation}
where $E_{\rm BPN}$ and $E_{\rm sub}$ are the total energies of isolated BPN and the metal substrate, respectively, while $E_{\rm BPN/sub}$ corresponds to the total energy of the BPN/metal(111) system. Here, $N_C$ is the number of C-atoms in the supercell. A more positive $E_{\rm ads}$ indicates stronger interaction, while a negative value implies that adsorption is thermodynamically unfavorable.
The $E_{\rm ads}$ for the different BPN/metal(111) configurations are presented in Table~S-I of the Supplementary Material.
Our calculations show that for Au(111), Ag(111), Al(111), Cu(111), and Ni(111), the variation in adsorption energy between different deposition sites is minimal, remaining below 1\%, suggesting relatively uniform interaction across these surfaces. However, for Pd(111) and Pt(111), we observe a site-dependent variation of approximately 13\%, which may be linked to the increased corrugation and local reorganization of the BPN layer upon interaction with these substrates. The case of Cu$_3$Au(111) lies in between, with adsorption energy variations of around 3\% between different configurations, indicating moderate site dependence due to 
its nature as a metallic alloy. 

 The $E_{\rm ads}$ for the most stable configurations are listed in Table~\ref{tab:strain}, our calculations reveal that BPN adsorption is exothermic across all metal surfaces except for Al(111), where $E_{\rm ads}$ is negative, indicating weaker interaction and reduced stability. The strongest interactions occur for Ni(111), followed by Pd(111), Pt(111), and Cu(111). In contrast, Au(111), Ag(111), and Cu$_3$Au(111) exhibit much lower adsorption energies, implying weaker physisorption-dominated interactions. 
 A clear correlation is observed between the adsorption energy and the equilibrium distance $\bar{h}$ between BPN and the metal surface. The metals that interact more strongly with BPN, such as Ni(111), Pd(111) and Pt(111), exhibit a significantly reduced interlayer distance, while weakly interacting substrates, such as Au(111) and Ag(111), result in a larger separation, characteristic of weak van der Waals interactions. This trend suggests that chemisorbed systems, which involve electronic hybridization between BPN and the substrate, tend to show stronger binding and structural distortions, whereas physisorbed systems largely preserve the intrinsic structure of BPN.
 
 Moreover, by comparing the adsorption energy trends with the degree of structural distortion, quantified through the out-of-plane corrugation parameter $\xi$, we observe that stronger interactions correlate with increased structural modifications. Weakly interacting metals such as Au, Ag, and Cu$_3$Au induce minimal corrugation ($\xi < 0.02$~\AA), allowing BPN to retain its nearly planar geometry. In contrast, strongly interacting metals such as Pd and Pt lead to significant corrugation ($\xi > 0.2$~\AA), as shown in Figs.~\ref{fig:struc}(b) and \ref{fig:struc}(c), which suggests enhanced charge redistribution and possible hybridization between electronic states of BPN and those of the metal substrate. Although, Ni and Cu present only slight structure distortions, with $\xi$ values of 0.04 and 0.03, respectively. Showing that even in strong interaction surfaces BPN can still show its characteristic planarity.  

To further understand the impact of the substrate on the BPN stability, we evaluate its formation energy ($E_{\rm for}$), defined as
\begin{equation}
    E_{\rm for} = (E_{\rm BPN/sub} - E_{\rm sub} - N_C\mu_C)/N_C,
\end{equation}
where $\mu_C$ represents the chemical potential of carbon, obtained from the energy per atom in graphite, used as the thermodynamic ground state for C. The formation energy indicates how the presence of a metal substrate modifies the energetic stability of BPN. For free-standing BPN, we compute $E_{\rm for} = 0.53$~eV/C-atom. As presented in Table~\ref{tab:strain}, when BPN is deposited onto metal substrates, we find that $E_{\rm for}$ consistently decreases, except for Al(111), suggesting that interaction with the metallic support stabilizes BPN. This effect is most pronounced for Ni(111), which shows a minimal BPN formation energy on top of that surface. Next, Pd(111) and Pt(111) present smaller values, while Al(111) provides the least stabilization, in agreement with its endothermic adsorption energy. 

\subsection{\label{sec:drho} Interfacial Charge Transfer and Bonding}

\begin{figure}
    \centering
    \includegraphics[width=1.0\linewidth]{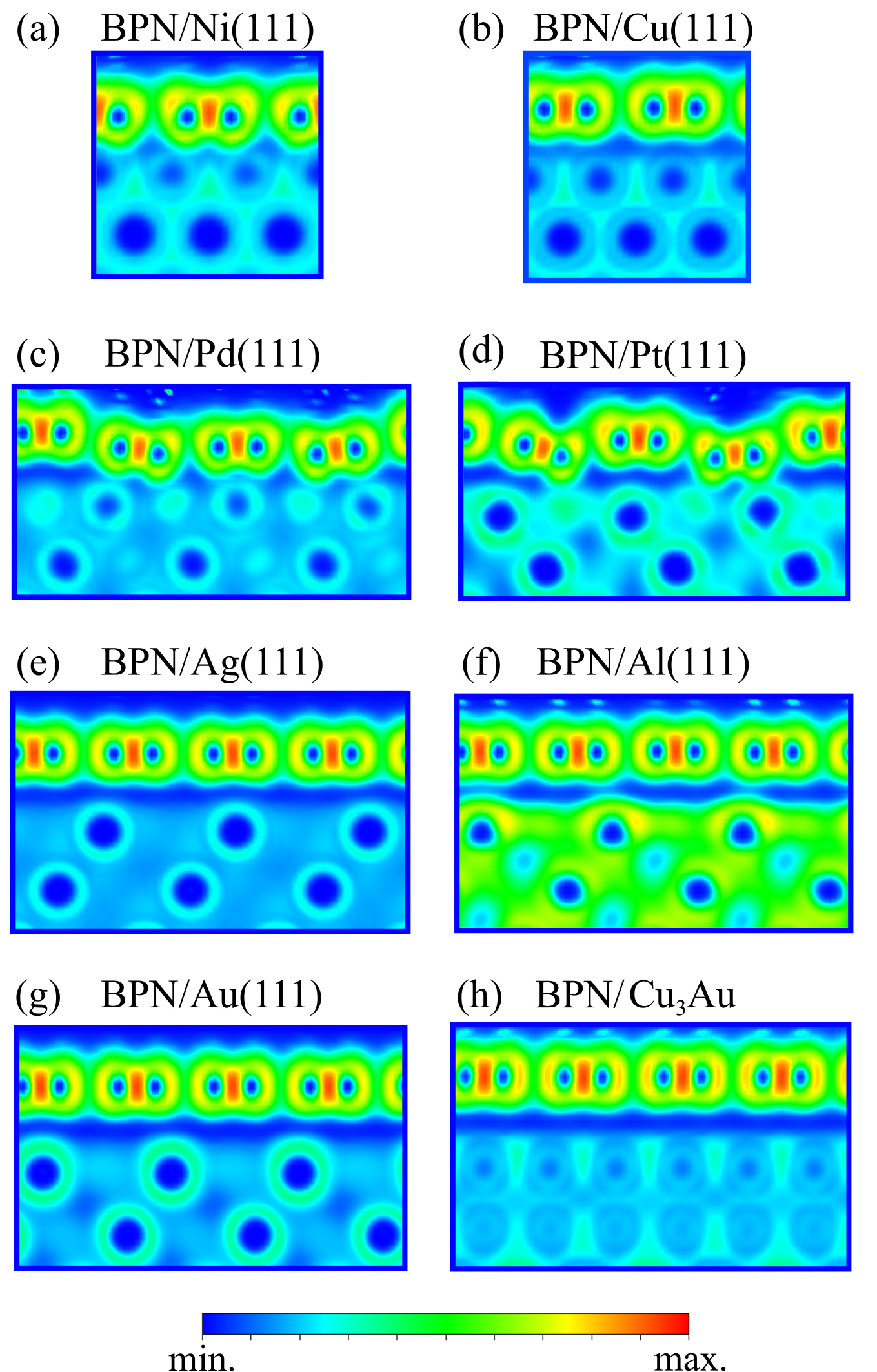}
    \caption{Electron localization function (ELF) maps for BPN adsorbed on the metal(111) surfaces investigated in this study. The subfigures correspond to: (a) BPN/Ni(111), (b) BPN/Cu(111), (c) BPN/Pd(111), (d) BPN/Pt(111), (e) BPN/Ag(111), (f) BPN/Al(111), (g) BPN/Au(111), and (h) BPN/Cu$_3$Au(111). Yellow to red regions denote high electron localization, while blue regions signify delocalized electronic states or metallic bonding character. The plane used for the ELF plots is represented by the dashed line in Fig.~\ref{fig:BFN}. }
    \label{fig:ELF}
\end{figure}

The charge transfer between BPN and the metal substrates was quantified using Bader charge analysis,\cite{Bader, Henkelman} with the results summarized in Table~\ref{tab:strain} as $\Delta\rho$. Positive values indicate electron donation from the metal to BPN, while negative values suggest charge depletion from the BPN layer.
We observe that the metal substrates act as electron donors, transferring charge to BPN, except for Au. The most substantial charge transfers occur for BPN/Ni(111) 
($\Delta\rho = 2.89\times 10^{14}$~e/cm$^{2}$) and BPN/Cu(111) ($\Delta\rho = 1.35\times 10^{14}$~e/cm$^{2}$), followed by BPN/Al(111) ($\Delta\rho = 1.09\times 10^{14}$~e/cm$^{2}$) and BPN/Pd(111) ($\Delta\rho = 1.04\times 10^{14}$~e/cm$^{2}$). These values reflect stronger metal–BPN interactions, consistent with their higher adsorption energies and the observed structural distortions. 
In the case of BPN/Pt(111), despite the moderate charge transfer
($\Delta\rho = 5.56\times 10^{13}$~e/cm$^{2}$), the significant distortion of BPN arises from strong local orbital hybridization, which induces chemical bonding without substantial net electron transfer.
Notably, the strong charge donation from Ni(111) correlates with a significant reduction in BPN formation energy and the emergence of spin polarization, indicating robust orbital hybridization and electronic coupling at the interface.
In contrast, noble metals such as Ag(111), Au(111), and also the alloy Cu$_3$Au(111) exhibit relatively low charge transfer values 
(6.64, –0.40, and 1.42~$\times 10^{13}$~e/cm$^{2}$, respectively), supporting the notion that their interactions with BPN are primarily governed by weak van der Waals forces. 

Interestingly, BPN/Au(111) is the only case with negative $\Delta\rho$, indicating a small net charge transfer from BPN to the gold surface. 
The computed work function (WF) of BPN is 4.33~eV, so this behavior aligns with the high work function of Au and its inert chemical character, suggesting that Au can slightly deplete electron density from BPN.
However, the observed charge transfer trend cannot be fully explained by the simple work function difference, since except for Al the WF of BPN is smaller than the metals (see WFs values in Table~\ref{tab:strain}). The direction and magnitude of charge transfer are strongly influenced by interfacial effects such as orbital hybridization, chemical bonding strength, and the formation of interface dipoles. 
\cite{hasegawa2011transfer,shao2019pseudodoping,khomyakov2009first}
These effects are particularly significant for strongly interacting substrates like Ni, Pd, and Pt, where electronic overlap and bond formation dominate over electrostatic considerations.

To further elucidate the nature of bonding at the interface, the Electron Localization Function (ELF) was computed, as shown in Fig.~\ref{fig:ELF}. 
The ELF highlights regions of localized electrons, helping differentiate between covalent, ionic, and metallic bonding. In the plots, red indicates high electron localization, typically associated with covalent bonds or lone pairs, while blue represents low localization, characteristic of delocalized or metallic bonding.
The ELF is plotted along a vertical plane passing through the C-bonds of the BPN square ring, as represented by the dashed line in Fig.~\ref{fig:BFN}.
In all systems, strong electron localization is observed within the BPN sheet, especially along the C–C bonds, consistent with its $\pi$-conjugated nature. The most notable differences arise in the ELF behavior at the interface region and within the metallic layers, revealing trends in interaction strength and bonding nature.
For Ni, Pd, and Pt (Figs.~\ref{fig:ELF}(a), \ref{fig:ELF}(c), and \ref{fig:ELF}(d), respectively), the ELF shows noticeable delocalization between the metal atoms and the adjacent carbon atoms. This suggests a moderate degree of covalent character, indicating stronger chemical interaction between these transition metals and the BPN sheet. The effect is most pronounced in the BPN/Ni(111) system, where the ELF contours near the interface are slightly more continuous, indicating enhanced orbital hybridization.  For Pd and Pt, the corrugation of the BPN sheet leads to variation in interaction strength; thus, carbon atoms located closer to the metal surface exhibit stronger bonding, while those farther away show weaker interaction.

In contrast, for Cu, Ag, and Au
(Figs.~\ref{fig:ELF}(b), \ref{fig:ELF}(e), and \ref{fig:ELF}(g), respectively), the ELF contours show a sharp transition at the interface with minimal overlap between the metal and carbon layers. The ELF remains largely localized on the carbon side, while the metal layer retains a diffuse metallic distribution. These features are consistent with a physisorption-type interaction, dominated by van der Waals (vdW) forces and weak orbital overlap. This is especially evident for Ag and Au, where the ELF at the interface is more symmetric and featureless. The BPN/Al(111) [Fig.~\ref{fig:ELF}(f)] interface exhibits a slightly elevated ELF in the metal layer compared to the noble metals, indicating a somewhat stronger interaction, potentially with partial charge transfer. However, the localization is still weaker than in the Ni, Pd, and Pt systems.
For the Cu$_3$Au alloy [Fig.~\ref{fig:ELF}(h)], the ELF profile resembles that of its constituents but with slightly more spatial modulation near the interface. This could point to a tunable interaction strength due to the mixed metallic environment, though the bonding remains predominantly non-covalent.

These results confirm the trend that substrates inducing stronger interactions (Ni, Pd, Pt) also facilitate more pronounced charge transfer to BPN, enhancing its electronic coupling and possibly modulating its catalytic and transport properties. On the other hand, weakly interacting metals such as Au and Ag preserve the intrinsic electronic features of BPN, which may be beneficial for applications requiring minimal substrate influence.

\subsection{Electronic Structure}

\begin{figure}
    \includegraphics[width=\columnwidth]{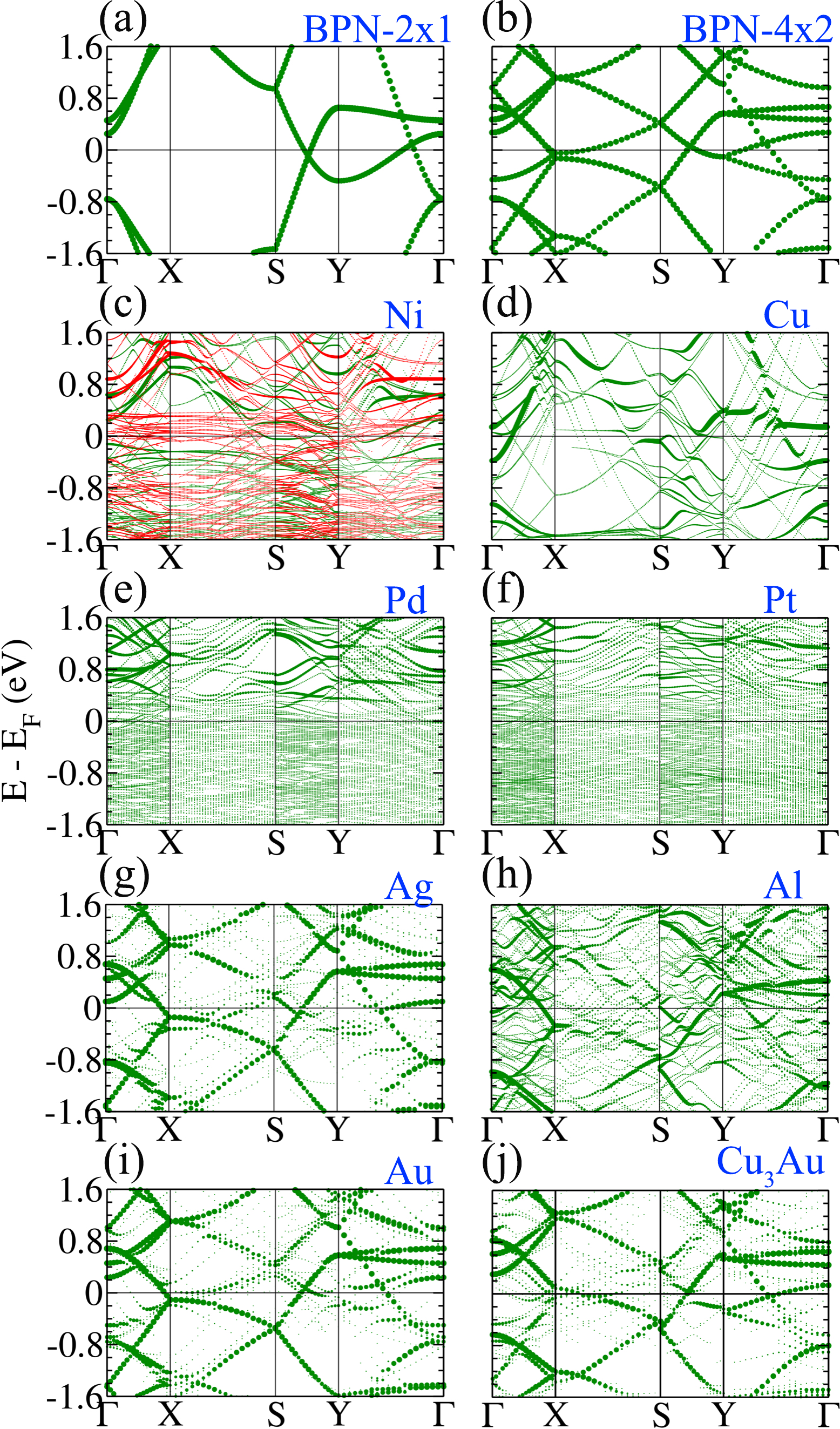}
    \caption{\label{fig:bands} Projected band structures of free-standing BPN with (a) $(2\times1)$ and (b) $(4\times2)$ periodicity. BPN-projected band structures for BPN deposited on (c) Ni(111), (d) Cu(111), (e) Pd(111), (f) Pt(111), (g) Ag(111), (h) Al(111), (i) Au(111), and (j) Cu$_3$Au(111). Note that the Ni(111) and Cu(111) systems adopt a $(2\times1)$ BPN supercell, while all other substrates use the $(4\times2)$ periodicity. In panel (c), green and red indicate spin-up and spin-down components, respectively.}
\end{figure}

\begin{figure*}
    \includegraphics[width=0.76\textwidth]{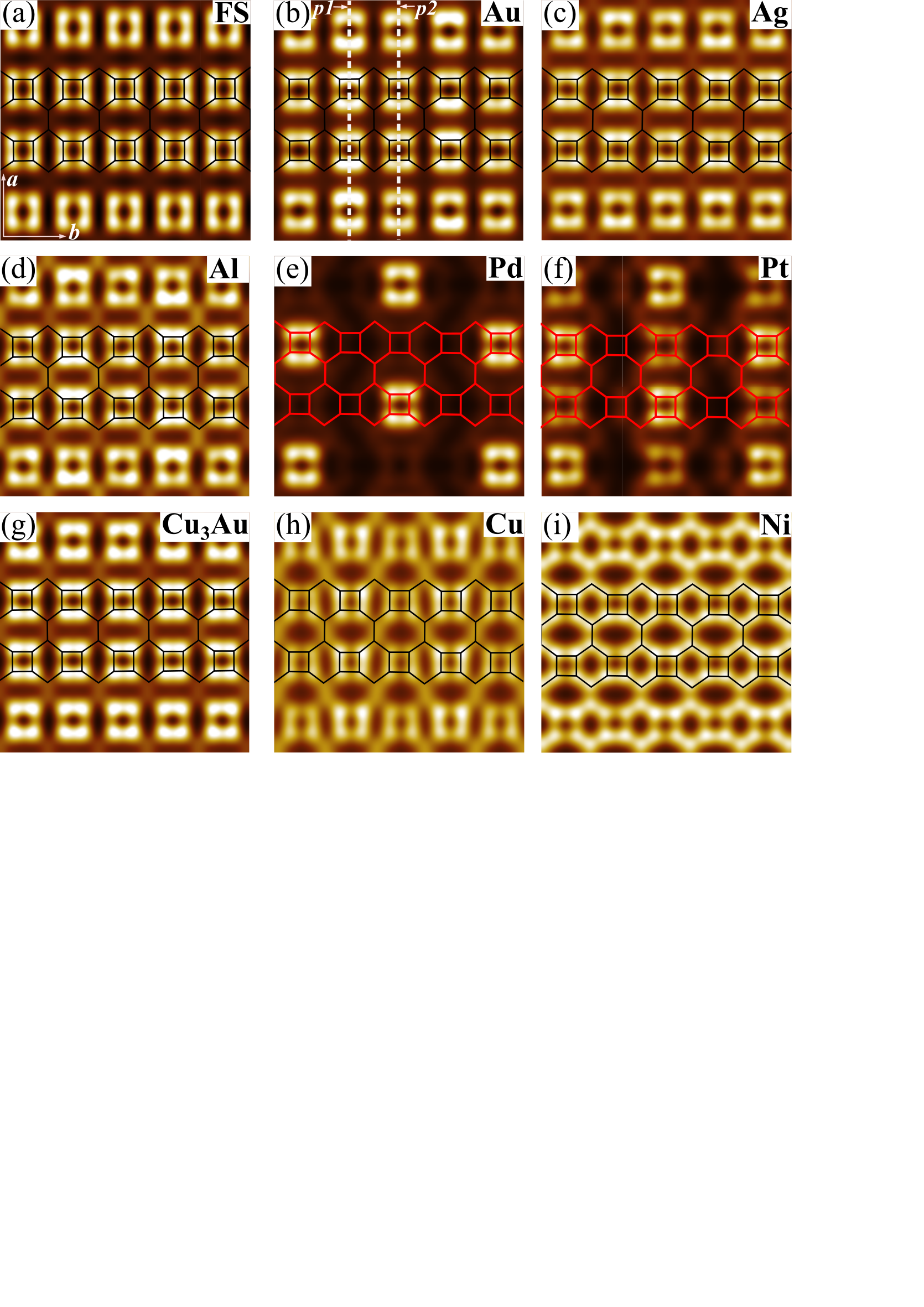}\vspace{-75mm}
    \caption{\label{fig:stm2}  Simulated constant-height Scanning Tunnelling Microscopy (STM) images of (a) the free‑standing biphenylene (BPN) and supported on a range of metal substrates of (b) Au, (c) Ag, (d) Al, (e) Pd, (f) Pt, (g) Cu$_3$Au, (h) Cu, and (i) Ni. The local density of states was integrated from the Fermi level up to E$_F+1.0$~eV (empty states) and sampled on a plane $\sim 1.60$~\AA~above the BPN, mimicking a constant‑height STM experiment. A common colour scale is used for all panels, running from 0 (dark) to $0.005~e$\AA$^{-3}$  (bright).}
\end{figure*}

To understand how metal substrates influence the electronic properties of BPN, we analyze the band structures of the BPN/metal(111) interfaces and compare them with the band structure of free-standing (FS) BPN. The degree of interaction between BPN and the substrate can be assessed by examining band hybridization, charge transfer effects, and modifications in the BPN projected band structures.
Fig.~\ref{fig:bands} displays the projection of BPN states onto the band structures of the investigated BPN/metal(111), alongside the band structures of free-standing BPN-($2\times 1$) [Fig.~\ref{fig:bands}(a)] and BPN-(4$\times$2) [Fig.~\ref{fig:bands}(b)] for comparison. The preservation of the BPN band structure on Ag, Au, and Cu$_3$Au indicates weak interactions between BPN and these substrates, consistent with physisorption mechanisms, as shown in Figs.~\ref{fig:bands}(g), \ref{fig:bands}(i), and \ref{fig:bands}(j). This weak interaction likely arises from the inert nature of these metals, which minimally disturb the electronic structure of BPN.
For Cu and Al [Figs.~\ref{fig:bands}(d) and \ref{fig:bands}(h)], the observed band hybridization points to stronger interactions, due to some level of charge transfer and formation of localized electronic states at the interface. 
These interactions suggest that Al and Cu influence the electronic properties of BPN more significantly, potentially altering its functionality in device applications.
In the case of Ni, Pd, and Pt the BPN band structures are significantly disrupted, reflecting the formation of covalent or chemical bonds, as discussed in Section~\ref{sec:drho}. 
This is evident in Figs.~\ref{fig:bands}(c), \ref{fig:bands}(e), and \ref{fig:bands}(f), which show pronounced deviations from the FS-BPN band structure. 
These metals are known for their higher chemical reactivity, which facilitates orbital overlap with BPN states, leading to strong coupling. Such strong interactions may result in substantial changes to the electronic properties of BPN. Furthermore, for BPN/Ni(111) the BPN states become spin polarized. We find a BPN magnetization of 0.05~$\mu_B$ per unit cell. A similar small magnetization has been reported for graphene deposited on Ni(111). \cite{zhang2014first,weser2010induced}

Overall, the nature of the interaction between BPN and the metal substrates varies significantly, ranging from weak physisorption to strong chemisorption. These findings highlight the importance of substrate selection in tailoring the electronic properties of BPN for specific applications, such as electronic devices, sensors, or catalysis.

Constant-height STM maps were simulated within the Tersoff–Hamann approximation \cite{tersoff1985theory} by integrating the LDOS from the Fermi level up to +1.0 eV on a plane 1.6~\AA~above the average BPN layer. Similar to the FS-BPN [Fig. \ref{fig:stm2}-(a)], for BPN adsorbed on Au(111) [Fig. \ref{fig:stm2}-(b)], Ag(111) [Fig. \ref{fig:stm2}-(c)] and Al(111) [Fig. \ref{fig:stm2}-(d)] the calculated images exhibit brightness confined to the four-membered carbon squares. This contrast is electronic in origin, an enhancement of the $\pi$-projected LDOS caused by weak overlap with substrate $s/p$ states, rather than topographic, since the surface corrugation is negligible ($\xi \leq$ 0.01~\AA). The behaviour accords with the ultraflat BPN islands reported experimentally on Au(111).\cite{fan2021biphenylene}

To elucidate the origin of the contrast we inspected the LDOS, evaluated over the same energy window, on two orthogonal vertical charge-density slices ($p1$ and $p2$, $xz$ plane) that intersect the interface. As shown in Figs. S-2-(a),(b), and (c), states localized on the selected top-layer metal atoms overlap with BPN $p_z$-orbitals; the overlap is strongest in the immediate interfacial region along the $p1$ plane, giving rise to the two distinct brightness patterns observed. The close similarity of the three STM patterns is therefore unsurprising, since all three systems share virtually (near) zero corrugation, comparable interlayer separations, and adsorption energies.

A marked change appears on Pd(111) [Fig. \ref{fig:stm2}-(e)], Pt(111) [Fig. \ref{fig:stm2}-(f)] and Cu(111) [Fig. \ref{fig:stm2}-(h)]. Here, the BPN lies closer to the metal (2.33, 2.59, and 2.52~\AA) and binds more strongly ($E_{ads}$ $=$ 0.11 - 0.17 eV). On Pt(111) and Pd(111) alternate four-membered squares form direct C–metal bonds that pull those rings $\xi =$ 0.30 Å (Pt) and $\xi = $ 0.25 Å (Pd) closer to the substrate, while the non-bonded squares remain higher. Because the STM signal is sampled on a constant-height plane 1.6~\AA~above the averaged BPN surface, only the geometrically elevated, non-bonded squares fall within the tunneling isosurface and therefore appear as the dominant protrusions in the empty-state map. On Cu(111) the backbone corrugates by only $\xi =$ 0.03~\AA, yet this subtle height modulation is enough to render every row of four-membered carbon squares slightly brighter than its neighbour, producing the faint row–by-row contrast observed in the simulated image.

The ordered Cu$_3$Au(111) alloy [Fig. \ref{fig:stm2}-(g)] represents an intermediate case: its inter-layer separation (3.23 ~\AA) and adsorption energy (0.09 eV) are comparable to those obtained for the noble-metal substrates, and the simulated STM map is almost featureless, indicating that every four-membered ring couples to the mixed Cu/Au surface layer in an essentially equivalent manner. By contrast, the Au(111), Ag(111) and Al(111) substrates exhibit row-by-row brightness modulations that trace the spatial variation of the interfacial LDOS documented in Fig. S-2. 

On Ni(111), the BPN [Fig. \ref{fig:stm2}-(i)] lies still closer to the substrate (1.94~\AA) and shows the strongest binding (E$_{ads}$ = 0.56 eV). Although the geometric corrugation is negligible ($\xi = $0.04~\AA), pronounced $\pi-d$ hybridisation enhances the empty-state tunnelling probability uniformly across all carbon sites. Hence, the constant-height STM image resolves the complete square-hexagon-octagon lattice rather than isolated bright squares. The progression from physisorption to strong chemisorption across the series thus explains the systematic evolution of the simulated STM contrast and underpins the experimental bias-dependent images observed for related non-benzenoid carbon networks.

\subsection{Biphenylene HER performance}

As global fossil fuel reserves continue to decline and environmental concerns become increasingly pressing, the transition to clean and sustainable energy sources has gained significant attention. Among the various renewable energy alternatives, hydrogen molecules (H$_2$) stand out as a promising candidate for a sustainable energy future.\cite{crabtree2008hydrogen} With its high energy density, zero emissions upon use, and broad applicability, hydrogen offers immense potential for addressing global energy demands. However, its storage remains a major challenge due to the weak intermolecular forces between H$_2$ molecules, which hinder efficient packing under ambient conditions.\cite{le2023fueling} Consequently, extensive research has been devoted to both hydrogen storage strategies and its production via the hydrogen evolution reaction (HER).\cite{lasia2019mechanism}

Two-dimensional (2D) materials have recently emerged as highly promising catalysts for HER, thanks to their distinctive electronic and structural properties.\cite{karmodak2020catalytic, li2021strategies} 
Recent theoretical studies have further highlighted the potential of BPN-based materials for HER applications. \cite{Luo2021}
For instance, boron–nitrogen co-doped BPN networks have been proposed as electronically and dynamically stable HER catalysts, with strain engineering and machine learning models predicting enhanced HER performance. \cite{yuan2024strain}
Additionally, defect engineering in metal-free BPN has been shown to activate its porous basal plane for HER, yielding a favorable adsorption energy of $\Delta G_{\rm H*} = 0.082$~eV.
\cite{sahoo2023activation}
Biphenylene nanoribbons (BPRs) also exhibit promising behavior, with electronic properties strongly dependent on ribbon width; notably, 15-BPR achieved $\Delta G_{\rm H*} = 0.005$~eV,\cite{somaiya2024biphenylene} outperforming even Pt(111) in HER activity.

In this work, we investigate the catalytic performance of BPN deposited on various metal(111) surfaces in acidic media for HER applications. Under such conditions, the reaction proceeds via the following fundamental steps:  
\begin{eqnarray}
    H^+ + e^- + * &\rightarrow& H*; \label{volmer} \\
    H* + H^+ + e^- &\rightarrow& H_2 + *; \label{heyrovsky} \\
    2H* &\rightarrow& H_2 \label{tafel}.
\end{eqnarray}
Here, Eq.~\ref{volmer} corresponds to the Volmer step, in which a proton ($H^+$) is reduced at an active site ($*$) on the catalyst surface. Molecular hydrogen (H$_2$) can then form via either the Heyrovsky step (Eq.~\ref{heyrovsky}), involving a second proton-electron transfer, or the Tafel step (Eq.~\ref{tafel}), where two adsorbed H atoms recombine. A key descriptor of HER activity is the Gibbs free energy of H adsorption ($\Delta G_{\rm H*}$), which is expressed as
\begin{equation}
    \label{eq:deltaG}
    \Delta G_{\rm H*} = \Delta E_{\rm H} + \Delta E_{\rm ZPE} - T\Delta S_{\rm H},
\end{equation}
where $\Delta E_{\rm H}$ represents the hydrogen adsorption energy, defined as $\Delta E_{\rm H} = E_{\rm sub+H} - E_{\rm sub} - \frac{1}{2}E_{\rm H_2}$, 
with $E_{\rm sub+H}$, $E_{\rm sub}$ and $E_{\rm H_2}$ denoting the total energies of the system with adsorbed H, the pristine catalyst, and an isolated H$_2$ molecule, respectively. The terms $\Delta E_{\rm ZPE}$ and $T\Delta S_{\rm H}$ account for zero-point energy and entropy contributions, which are generally independent of the catalytic material.\cite{norskov2005trends, zhou2019transition} Based on established values from Ref.\citenum{norskov2005trends}, widely used in HER studies on 2D materials,\cite{sahoo2023activation, qu2018effect, sajjad2023colossal, qu2015first, zhu2019single} Eq.~\ref{eq:deltaG} simplifies to $\Delta G_{\rm H*} = \Delta E_{\rm H} + 0.24$~eV. This provides a reliable framework for assessing the HER catalytic performance of BPN/metals(111). For all systems, we have considered a low hydrogen coverage of 1/48, corresponding to one atomic H in a BPN-($4\times 2$) supercell. 
\begin{table}
\caption{\label{tab:HER} Gibbs free energy of hydrogen adsorption ($\Delta G_{\rm H*}$) for free-standing biphenylene (FS-BPN) and for BPN supported on various metal(111) substrates, evaluated at two distinct adsorption sites (C1 and C2, see Fig.~\ref{fig:BFN}). Values in parentheses for FS-BPN are taken from Ref.~\onlinecite{sahoo2023activation}.}
\begin{ruledtabular}
\begin{tabular}{ccc}
metal(111) & \multicolumn{2}{c}{$\Delta G_{\rm H*}$~(eV)}  \\
substrate  & site C1 & site C2  \\
\hline
FS-BPN & 0.13 (0.08) & 1.25 (1.17)  \\
Ag & 0.07 & 1.12  \\
Al & 0.31 & 0.25  \\
Au & 0.11 & 1.11  \\
Cu$_3$Au & 0.14 & 1.14 \\
Cu & -0.09 & 0.49 \\
Ni & -0.11 & 0.24 \\
Pd &  0.04 & 0.28 \\
Pt &  0.07 & 0.38 \\
\end{tabular}
\end{ruledtabular}
\end{table}

The results presented in Table~\ref{tab:HER} highlight the variation in Gibbs free energy of H adsorption ($\Delta G_{\rm H*}$) across different metal(111) surfaces. This parameter is crucial for evaluating HER performance, as an ideal catalyst should exhibit a $\Delta G_{\rm H*}$ value close to zero, ensuring optimal H adsorption and desorption kinetics. The $\Delta G_{\rm H*}$ for FS-BPN  is also presented.
BPN structure contains two symmetry-inequivalent C-atoms in its unit cell, represented by sites C1 and C2 in Fig.~\ref{fig:BFN}. Site~C1 (0.13 eV) presents a much lower HER activity than C2 (1.25 eV), in agreement with previous DFT works. \cite{sahoo2023activation}
For site~C1, the $\Delta G_{\rm H^*}$ values in BPN/metal(111) systems are mostly lower or similar to that of FS-BPN, with absolute values ranging from 0.04--0.14~eV. The exception is for BPN/Al(111), which shows a considerably higher value of 0.31~eV. These values can be evaluated in the context of the optimal thermoneutral range, typically considered to be between $-0.1$ and $+0.1$~eV, for efficient HER activity. 
\cite{santos2011hydrogen}



\begin{figure}
    \centering
    \includegraphics[width=0.8\linewidth]{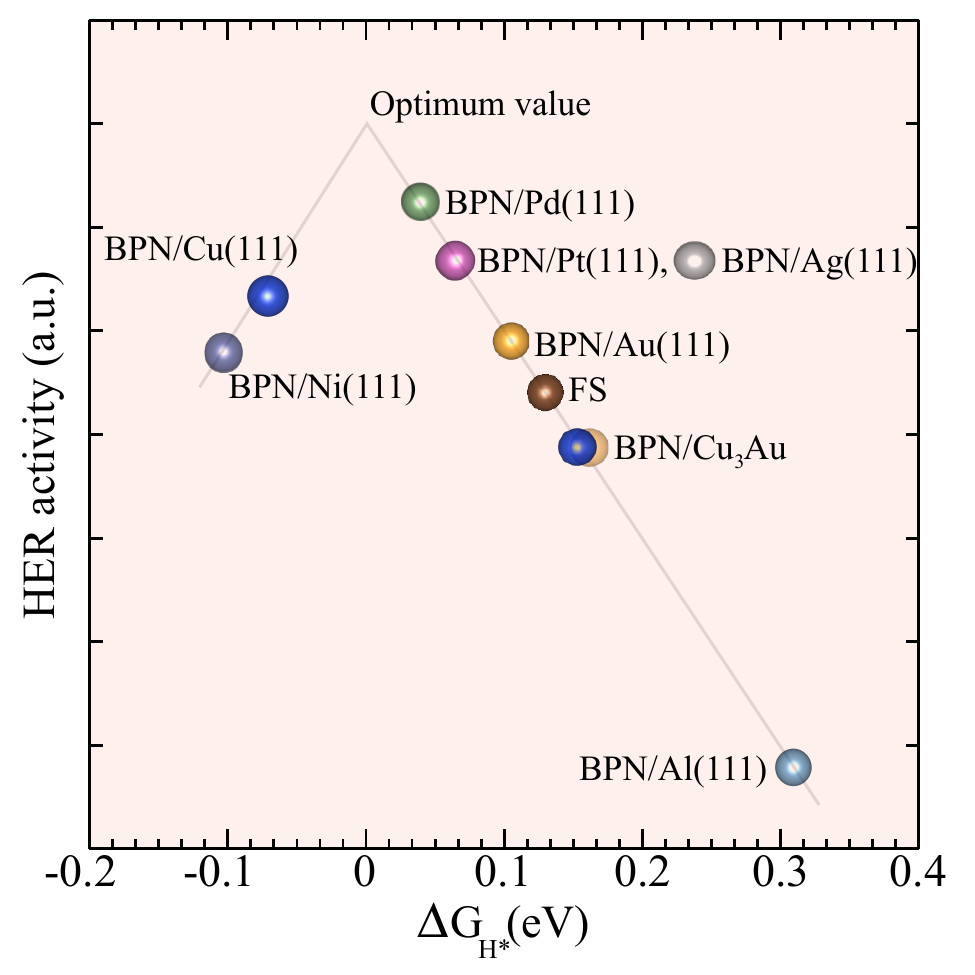}
    \caption{Volcano plot illustrating the HER activity of BPN supported on Ni(111) (dark gray sphere), Cu(111) (blue sphere), Pd(111) (green sphere), Pt(111) (magenta sphere), Ag(111) (light gray sphere), Au(111) (dark yellow sphere), and Cu$_3$Au(111) (dark blue and yellow spheres), along with freestanding BPN (brown sphere). The peak of the volcano plot corresponds to the optimal $\Delta G_{\rm H*}$ value for HER activity.}
    \label{fig:volcano}
\end{figure}

A volcano plot for site C1 (Fig.~\ref{fig:volcano}) was constructed to directly compare the HER activity of the investigated systems. The peak of the plot corresponds to the optimal $\Delta G_{\rm H*}\sim 0$ value, representing the most favorable condition for efficient H evolution. The calculated 
    $\Delta G_{\rm H*}$ values for BPN on various metal(111) substrates span both below and above the optimal value; consequently, the corresponding HER activities appear on both sides of the volcano peak in Fig.~\ref{fig:volcano}. BPN supported on Pt(111), Pd(111), and Ag(111) lies closest to the volcano apex, indicating near-optimal HER performance. In contrast, BPN/Ni(111) and BPN/Cu(111), located on the left side of the peak (with negative $\Delta G_{\rm H*}$), suggest stronger H binding, which may hinder desorption and reduce overall catalytic efficiency. On the right side, BPN/Au(111), BPN/Cu$_3$Au(111), and FS-BPN exhibit positive $\Delta G_{\rm H*}$ values, reflecting weaker H adsorption and potential limitations in proton capture. BPN/Al(111), with the highest $\Delta G_{\rm H*}$ value among the studied systems, is positioned far from the optimal region, indicating poor HER activity. 

For H adsorption at site C2, $\Delta G_{\rm H*}$ decreases for all BPN structures supported on metal(111) surfaces compared to FS-BPN. While the reduction is approximately 10\% ($\sim$1.1~eV) for Ag, Au, and Cu$_3$Au, it exceeds 60\% (0.24–0.49~eV) for the other substrates. 
An important observation from Table~\ref{tab:HER} is the large difference in $\Delta G_{\rm H*}$ between sites C1 and C2 for FS-BPN, indicating site-dependent variations in H adsorption. This discrepancy suggests that intrinsic structural or electronic inhomogeneities within BPN play a significant role in its catalytic behavior. 
However, when BPN is supported on certain metal substrates, this difference decreases, implying that the metal support helps homogenize the adsorption properties across different sites. This effect can be attributed to the interaction between BPN and the metal, which modifies the local electronic environment and reduces site-specific variations in H binding. 
The ability of metal substrates to stabilize and equalize adsorption sites further underscores their influence on the catalytic performance of BPN-based systems.

In summary, the HER activity of BPN is highly dependent on the choice of metal substrate. While noble metals such as Pt, Pd, and Ag offer superior catalytic performance, alternative substrates like Ni and Cu may provide a promising balance between performance and cost-effectiveness. These insights emphasize the importance of substrate selection in optimizing BPN-based HER catalysts and guide future experimental efforts in developing efficient and sustainable hydrogen production technologies.

\section{Conclusions}

Using first-principles calculations, we have investigated the structural, electronic, and catalytic properties of biphenylene supported on various metal(111) surfaces. We showed that the interaction strength between biphenylene and the metal substrates ranges from physisorption (e.g., Au, Ag) to chemisorption (e.g., Ni, Pd, Pt), directly influencing the structural corrugation, electronic hybridization, and charge transfer at the interface. For strongly interacting metals, significant band structure modifications and, in the case of Ni(111), spin polarization were observed. Furthermore, we evaluated the hydrogen evolution reaction (HER) activity and demonstrated that certain metal supports, especially Pt(111), Pd(111), Ag(111), Cu(111), and Ni(111), substantially enhance the catalytic performance of biphenylene. These findings reveal the potential of metal-supported biphenylene systems for applications in nanoelectronics and sustainable hydrogen production technologies.

\begin{acknowledgments}

The authors acknowledge financial support from FAPEMIG and CAPES, and computer time from LCC-UFLA and CENAPAD-SP.

\end{acknowledgments}


\begin{thebibliography}{57}
\bibitem [{str()}]{straineq}%
  \BibitemOpen
  \href@noop {} {}\bibinfo {note} {The lattice mismatch along the $x$ and $y$
  directions are defined, respectively, by $\epsilon_x = [(m a_{\rm BPN} -
  pa_{\rm lat})/(pa_{\rm lat})]\times 100\%$, $\epsilon_y = [(n b_{\rm BPN} -
  qa_{\rm lat})/(qa_{\rm lat})]\times 100\%$. And the average lattice mismatch
  by $\overline{\epsilon} = (|\epsilon_x| + |\epsilon_y|)/2$.}\BibitemShut
  {Stop}%
\end{thebibliography}

\providecommand{\noopsort}[1]{}\providecommand{\singleletter}[1]{#1}%

\end{document}